\documentclass[prd,aps,floatfix,nofootinbib,11 pt]{revtex4}
\usepackage{amsmath,graphicx,color,epsfig}

\input{tcilatex}

\begin{document}

\title{Decoherence dynamics of geometric measure of quantum discord and
measurement induced nonlocality for noninertial observers at finite
temperature}
\author{M. Ramzan\thanks{%
mramzan@phys.qau.edu.pk}}
\address{Department of Physics Quaid-i-Azam University \\
Islamabad 45320, Pakistan}
\date{\today }

\begin{abstract}
Quantum discord quantifies the total non-classical correlations in mixed
states. It is the difference between total correlation, measured by quantum
mutual information, and the classical correlation. Another step forward
towards the quantification of quantum discord was by Dakic, Vedral, and
Brukner [Phys. Rev. Lett. \textbf{105},190502 (2010)] who introduced the
geometric measure of quantum discord (GMQD) and derived an explicit formula
for a two-qubit state. Recently, Luo and Fu [Phys. Rev. Lett. \textbf{106},
120401 (2011)] introduced measurement-induced nonlocality (MIN) as a measure
of nonlocality for a bipartite quantum system. The dynamics of GMQD is
recently considered by Song et al. [arXiv: quant/ph.1203.3356] and Zhang et
al. [Eur. Phys. J. D \textbf{66,} 34 (2012)] for inertial observers.
However, the topic requires due attention in noninertial frames,
particularly, from the perspective of MIN. Here I consider $X$-structured
bipartite quantum system in noninertial frames and analyze the decoherence
dynamics of GMQD and MIN at finite temperature. The dynamics under the
influence of amplitude damping, depolarizing and phase flip channels is
discussed. It is worth-noting that initial state entanglement plays an
important role in bipartite states. It is possible to distinguish the Bell,
Werner and general type initial quantum states using GMQD. Sudden transition
in the behaviour of GMQD and MIN occurs depending upon the mean photon
number of the local environment. The transition behaviour disappears for
larger values of $\bar{n},$ i.e. $\bar{n}>0.3.$ It becomes more prominent,
when environmental noise is introduced in the system. In the presence of
environmental noise, as we increase the value of acceleration $r$, GMQD and
MIN decay due to Unruh effect. The effect is prominent for the phase flip
and amplitude damping channels. However, in case of depolarizing channel, no
sudden change in the behaviour of GMQD and MIN is observed. The
environmental noise has stronger affect on the dynamics of GMQD and MIN as
compared to the Unruh effect. Furthermore, Werner like states are more
robust than General type initial states at finite temperature.\newline
\end{abstract}

\maketitle

\address{Department of Physics Quaid-i-Azam University \\
Islamabad 45320, Pakistan}


\vspace*{1.0cm}

\vspace*{1.0cm}

%\date{\today}

%\newpage

\section{Introduction}

Quantum entanglement, a special quantum correlation, has attracted
considerable attention during recent years because many quantum information
processes depend on entanglement [1]. Main difficulty with entanglement is
its quantification. Different entanglement measures have been proposed [2,
3]. Quantum correlation, one of the key features in quantum information
theory, has become an important tool to study quantum many-body systems, for
example, quantum phase transition in different correlated systems. However,
the quantitative and qualitative evaluation of such correlations remains an
open problem. So far, several quantifiers of non-classicality of
correlations have been introduced in literature [4-10], but still there is
no clear criteria for the faithfulness of them. Moreover, for the
multipartite states, the geometric measure of entanglement (GME) has been
proposed [11-16]. The topic attracted due respect in short span of time and
bounds on several entanglement measures like von Neumann Entropy [17],
Negativity [18], Concurrence [19] and GME [20, 21] were obtained.

Recent investigations reveal that there exist quantum correlations other
than entanglement. Quantum discord [22, 23] quantifies the total
non-classical correlations in a quantum state. It was suggested that the
quantum discord, rather than entanglement, is responsible for the efficiency
of a quantum computer, which is confirmed both theoretically [24] and
experimentally [25]. A systematical analysis of quantum and classical
correlations for bipartite and multipartite quantum systems have been
proposed by Okrasa and Walczak [26]. Quantum discord for the subclass of
so-called $X$--states [27-30], a qubit-qutrit [31] and qubit-qudit [32]
systems have also been proposed. Huang et al. [33] have proposed a new
criterion for judging zero quantum discord for arbitrary bipartite states.
Yao et al. [34] have proposed the geometric interpretation of the geometric
discord. The dynamics of quantum discord for a two-qubit system in a quantum
spin environment have been proposed by Guo et al. [35]. Recently, discord
for multipartite quantum states has been investigated [36-37] and its
dynamics under decoherence [38].

However, the quantum discord for a general two qubit state remains a
nontrivial task and only the lower and upper bounds were investigated [39,
40]. Motivated from the difficultly in computing the quantum discord, the
geometric measure of quantum discord (GMQD) was proposed [6], which
quantifies the amount of non-classical correlations of a state in terms of
its minimal distance from the set of genuinely classical states. It can be
defined as the nearest distance between the given state and the set of
zero-discord states. Recently, Luo and Fu [41, 42] have introduced a
geometric measure of nonlocality which they termed as measurement-induced
nonlocality (MIN). It can be defined as the maximum distance between the
bipartite state and its post-measurement state, where the maximum is taken
over all the von Neumann local measurements which do not disturb the local
state. Recent investigations in this direction includes [43-60], where the
behaviour of the dynamics of the system is discussed under different
scenarios. Furthermore, the dynamics of entanglement at zero and finite
temperature has been studied by many authors, for example [61-63].

The most difficult problem in realizing the quantum information technology
is that the quantum system can never be isolated from the surrounding
environment completely. Interactions with the environment deteriorate the
purity of the quantum states. This general phenomenon, known as decoherence
[64], is a serious obstacle against the preservation of quantum
super-positions over long periods of time. Decoherence entails non-unitary
evolutions, with serious consequences, like a loss of information and
probable leakage toward the environment. Therefore, in a realistic and
practical situation, decoherence caused by an external environment is
inevitable and the influence of an external environmental system on the
entanglement cannot be ignored. However in some cases it can create quantum
correlations in the system [65]. Understanding the dynamics of open quantum
systems is of considerable importance. The Schrodinger equation, which
describes the evolution of closed systems, is generally inapplicable to open
systems, unless one includes the environment in the description. This is,
however, generally difficult, due to the large number of environment degrees
of freedom. An alternative is to develop a description for the evolution of
only the subsystem of interest. Much attention has been given to the
phenomenon of decoherence that causes an irreversible transfer of
information from the system to the environment [66, 67] with the special
attention to the degradation of entanglement [68-73]. On the other hand, the
behaviour of entanglement in noninertial frames was investigated for the
first time by Alsing et al. [74]. The subject have attracted much attention
during recent years [75-93]. It has also been investigated under decoherence
for qubit-qubit [94-97], qubit-qutrit system [98] and multipartite systems
[99]. The entanglement dynamics for noninertial observers in a correlated
environment is considered in Ref. [100], where it is shown that correlated
noise compensates the loss of entanglement caused by the Unruh effect.
Recently, Oliveira et al. [101] have studied the entanglement measure for
pure six-qubit quantum states. Whereas Zehua and Jiliang [102] have studied
how the Unruh effect affects the transition between classical and quantum
decoherence for a general class of initial states.

In this paper, I have investigated the decoherence dynamics of geometric
measure of quantum discord and measurement-induced nonlocality at finite
temperature for $X$-type initial states in relativistic frames. The two
observers Alice and Bob share an $X$-type state in noninertial frames. Alice
is considered to be stationary whereas Bob moves with a uniform acceleration
$r$. It is shown that initial state entanglement plays an important role in
bipartite quantum states. Different decoherence channels are considered
parameterized by decoherence parameter $p$ such that $p\in \lbrack 0,1]$.
The lower and upper limits of decoherence parameter represent the fully
coherent and fully decohered system, respectively. Whereas, the lower and
upper limits of parameter $X$ correspond to $t=\infty ,$ $0$, respectively.
It is seen that different initial states can be distinguished using GMQD
such as the Bell diagonal, Werner and general type initial states. It is
also seen that the depolarizing channels heavily influences the dynamics of
GMQD and MIN as compared to the amplitude damping channel. Furthermore, no
GMQD and MIN sudden death is seen at finite temperature even in the presence
of decoherence.

\section{Decoherence dynamics of accelerated observers}

The evolution of a system and its environment can be described by
\begin{equation}
U_{SE}(\rho _{S}\otimes |0\rangle _{E}\left\langle 0\right\vert
)U_{SE}^{\dag }
\end{equation}%
where $U_{SE}$ represents the evolution operator for the combined system and
$|0\rangle _{E}$ corresponds to the initial state of the environment. By
taking trace over the environmental degrees of freedom, the evolution of the
system can be obtained as%
\begin{eqnarray}
L(\rho _{S}) &=&\text{Tr}_{E}\{U_{SE}(\rho _{S}\otimes |0\rangle
_{E}\left\langle 0\right\vert )U_{SE}^{\dag }\}  \notag \\
&=&\sum\nolimits_{\mu }\ _{E}\left\langle \mu \right\vert U_{SE}|0\rangle
_{E}\rho _{SE}\left\langle 0\right\vert )U_{SE}^{\dag }|\mu \rangle _{E}
\end{eqnarray}%
where $|\mu \rangle _{E}$ represents the orthogonal basis of the environment
and $L$ is the operator describing the evolution of the system. The above
equation can also be written as%
\begin{equation}
L(\rho _{S})=\sum\nolimits_{\mu }M_{\mu }\rho _{S}M_{\mu }^{\dag }
\end{equation}%
where $M_{\mu }=\ _{E}\left\langle \mu \right\vert U_{SE}|0\rangle _{E}$ are
the Kraus operators as given in Ref. [103]. The Kraus operators satisfy the
completeness relation%
\begin{equation}
\sum\nolimits_{\mu }M_{\mu }^{\dag }M_{\mu }=1
\end{equation}%
The decoherence process can also be represented by a map in terms of the
complete system-environment state. The dynamics of a $d$-dimensional quantum
system can be represented by the following map [104]%
\begin{equation}
U_{SE}|\xi _{l}\rangle _{S}\otimes |0\rangle _{E}=\sum\nolimits_{k}M_{k}|\xi
_{l}\rangle _{S}\otimes |k\rangle _{E}
\end{equation}%
where $\{|\xi _{l}\rangle _{S}\}$ $(l=1,.....,d)$ is the complete basis for
the system and%
\begin{eqnarray}
|\xi _{1}\rangle _{S}\otimes |0\rangle _{E} &\rightarrow &M_{0}|\xi
_{1}\rangle _{S}\otimes |0\rangle _{E}+.....+M_{d^{2}-1}|\xi _{1}\rangle
_{S}\otimes |d^{2}-1\rangle _{E}  \notag \\
|\xi _{2}\rangle _{S}\otimes |0\rangle _{E} &\rightarrow &M_{0}|\xi
_{2}\rangle _{S}\otimes |0\rangle _{E}+.....+M_{d^{2}-1}|\xi _{2}\rangle
_{S}\otimes |d^{2}-1\rangle _{E}  \notag \\
. &&  \notag \\
. &&  \notag \\
. &&  \notag \\
|\xi _{d}\rangle _{S}\otimes |0\rangle _{E} &\rightarrow &M_{0}|\xi
_{d}\rangle _{S}\otimes |0\rangle _{E}+.....+M_{d^{2}-1}|\xi _{d}\rangle
_{S}\otimes |d^{2}-1\rangle _{E}
\end{eqnarray}

Let Alice and Bob (the accelerated observer) share the following $X$-type
initial state [105]%
\begin{equation}
\rho _{AB}=\frac{1}{4}\left( I_{AB}+\sum\limits_{i=1}^{3}c_{i}\sigma
_{i}^{(A)}\otimes \sigma _{i}^{(B)}\right)
\end{equation}%
where $I_{AB}$ is the identity operator in a two-qubit Hilbert space, $%
\sigma _{i}^{(A)}$ and $\sigma _{i}^{(B)}$ are the Pauli operators of the
Alice's and Bob's qubit and $c_{i}$ $(0\leq |c_{i}|\leq 1)$ are real numbers
satisfying the unit trace and positivity conditions of the density operator $%
\rho _{AB}.$ In order to study the entanglement dynamics, different cases
for initial state are considered, for example, the general initial state $%
(|c_{1}|=0.7,|c_{2}|=0.9,|c_{3}|=0.4)$, the Werner initial state $%
(|c_{1}|=|c_{2}|=|c_{3}|=0.8)$, and Bell basis state $%
(|c_{1}|=|c_{2}|=|c_{3}|=1)$.

Let the Dirac fields, as shown in Refs. [83, 84], from an inertial
perspective, can be described by a superposition of Unruh monochromatic
modes $|0_{U}\rangle =\otimes _{\omega }|0_{\omega }\rangle _{U}$ and $%
|1_{U}\rangle =\otimes _{\omega }|1_{\omega }\rangle _{U}$ with
\begin{equation}
|0_{\omega }\rangle _{U}=\cos r|0_{\omega }\rangle _{I}|0_{\omega }\rangle
_{II}+\sin r|1_{\omega }\rangle _{I}|1_{\omega }\rangle _{II}
\end{equation}%
and%
\begin{equation}
|1_{\omega }\rangle _{M}=|1_{\omega }\rangle _{I}|0_{\omega }\rangle _{II}
\end{equation}%
where $\cos r=(e^{-2\pi \omega c/a}+1)^{-1/2}$, $a$ is the acceleration of
the observer, $\omega $ is frequency of the Dirac particle and $c$ is the
speed of light in vacuum. The subscripts $I$ and $II$ of the kets represent
the Rindler modes in region $I$ and $II$, respectively, as shown in the
Rindler spacetime diagram (see Ref. [92], Fig. (1)). By using equations (8)
and (9), equation (7) can be re-written in terms of Minkowski modes for
Alice ($A$) and Rindler modes for Bob $(\tilde{B})$. The single-mode
approximation is used in this study, i.e. a plane wave Minkowski mode is
assumed to be the same as a plane wave Unruh mode (superposition of
Minkowski plane waves with single-mode transformation to Rindler modes).
Therefore, Alice being an inertial observer while her partner Bob who is in
uniform acceleration, are considered to carry their detectors sensitive to
the $\omega $ mode. To study the entanglement in the state from their
perspective one must transform the Unruh modes to Rindler modes. Hence,
Unruh states must be transformed into the Rindler basis. Let Bob detects a
single Unruh mode and Alice detects a monochromatic Minkowski mode of the
Dirac field. Considering that an accelerated observer in Rindler region $I$
has no access to the field modes in the causally disconnected region $II$
and by taking the trace over the inaccessible modes, one obtains the
following density matrix%
\begin{eqnarray}
\rho _{A\tilde{B}} &=&\frac{1}{4}\left(
\begin{array}{cccc}
(1+c_{3})\cos ^{2}r & 0 &  &  \\
0 & (1+c_{3})\sin ^{2}r+(1-c_{3}) &  &  \\
0 & c^{+}\cos r &  &  \\
c^{-}\cos r & 0 &  &
\end{array}%
\right.  \notag \\
&&\left.
\begin{array}{cccc}
&  & 0 & c^{-}\cos r \\
&  & c^{+}\cos r & 0 \\
&  & (1-c_{3})\cos ^{2}r & 0 \\
&  & 0 & (1+c_{3})+(1-c_{3})\sin ^{2}r%
\end{array}%
\right)
\end{eqnarray}%
where $c^{+}=c_{1}+c_{2}$ and $c^{-}=c_{1}-c_{2}.$

Since noise is a major hurdle while transmitting quantum information from
one party to other through classical and quantum channels. This noise causes
a distortion of the information sent through the channel. It is considered
that the system is strongly correlated quantum system, the correlation of
which results from the memory of the channel itself. The action of a two
qubit Pauli channel when both the qubits of Alice and Bob are streamed
through it, can be described in operator sum representation as [1]
\begin{equation}
\rho _{f}=\sum\limits_{k_{1},\text{ }k_{2}=0}^{1}(A_{k_{2}}\otimes
A_{k_{1}})\rho _{in}(A_{k_{1}}^{\dagger }\otimes A_{k_{2}}^{\dagger })
\end{equation}%
where $\rho _{in}$ represents the initial density matrix for quantum state
and $A_{k_{i}}$\ are the Kraus operators. A detailed list of single qubit
Kraus operators for different quantum channels under consideration is given
in table 1. In order to quantify the quantum correlations, the dynamics of
the system-environment interaction is investigated and only the reduced
matrices are considered. It is assumed that both Alice and Bob's qubits are
influenced by the environment. The reduced-density matrix of the inertial
subsystem $A$ and the noninertial subsystem $\tilde{B}$, can be obtained by
taking the partial trace of $\rho _{A\tilde{B}E_{A}E_{\tilde{R}}}=\rho _{A%
\tilde{B}}\otimes \rho _{A\tilde{B}E_{A}E_{\tilde{B}}}$\ over the degrees of
freedom of the environment i.e.
\begin{equation}
\rho _{A\tilde{B}}=\text{Tr}_{E_{A}E_{\tilde{B}}}\{\rho _{A\tilde{B}E_{A}E_{%
\tilde{B}}}\}.
\end{equation}

\section{GMQD and MIN at finite temperature}

Let us consider a two-qubit system (two two-level atoms) is interacting with
a thermal reservoir. Unlike the Ref. [103], where the authors studied the
system at $T=0$, in the present scheme, the effect of heat is included. The
dynamics of the density matrix $\hat{\rho}$ describing the two qubit system
is given by [106]%
\begin{equation}
\frac{d\rho }{dt}=\frac{1}{2}(\bar{n}+1)\Gamma \sum\limits_{i=1}^{2}\left\{ %
\left[ \sigma _{-}^{i},\rho \sigma _{+}^{i}\right] +\left[ \sigma
_{-}^{i}\rho ,\sigma _{+}^{i}\right] \right\} +\frac{1}{2}\bar{n}\Gamma
\sum\limits_{i=1}^{2}\left\{ \left[ \sigma _{+}^{i},\rho \sigma _{-}^{i}%
\right] +\left[ \sigma _{+}^{i}\rho ,\sigma _{-}^{i}\right] \right\}
\end{equation}%
where $\Gamma $ is the spontaneous emission rate, $\sigma _{\pm }^{i}$ $%
(i=1;2)$ are the rasing $(+)$ and lowering $(-)$ operators of atom $i$
defined as $\sigma _{+}^{i}=\left\vert 1\right\rangle \left\langle
0\right\vert _{i}$, $\sigma _{-}^{i}=\left\vert 0\right\rangle \left\langle
1\right\vert _{i},$ $\bar{n}$ is the mean occupation number of the reservoir
(assumed to be the same for both qubits) that can be related to the
parameter $X=\exp ^{-\Gamma (2\bar{n}+1)t}.$ Here, the quantity $X$ is the
time-dependent parameter which can be used to monitor the evolution of GMQD
and MIN for the the two-qubit system. Notice that at $t=0$, $X=1$, and that
at $t=\infty $, $X=0$. Therefore, physically meaningful values for $X$ are
between $0$ and $1$. On the right hand side of equation (13), the first term
describes the depopulation of the atoms due to stimulated and spontaneous
emission, while the second term corresponds to the re-excitations caused by
the finite temperature. The final state of the system (for the initial state
of the form equation (10)) keeps its $X$-form at any time during its
evolution and leads to
\begin{equation}
\left[
\begin{array}{cccc}
\rho _{11}[t] & 0 & 0 & \rho _{141}[t] \\
0 & \rho _{22}[t] & \rho _{23}[t] & 0 \\
0 & \rho _{32}[t] & \rho _{33}[t] & 0 \\
\rho _{41}[t] & 0 & 0 & \rho _{44}[t]%
\end{array}%
\right]
\end{equation}%
The master equation of the system, governed by the first-order coupled
differential equations, can be solved to yield the expressions of the final
density matrix as given in appendix-A [107]. The quantum discord, a measure
of the minimal loss of correlation in the sense of quantum mutual
information, can be defined for a bipartite quantum state as [22]%
\begin{equation}
D_{A}(\rho ):=\underset{\Pi ^{A}}{\min }\{I(\rho )-I(\rho |\Pi ^{A})\}
\end{equation}%
where minimum is taken over all local von Neumann measurements $\Pi ^{A}.$%
\begin{equation}
I(\rho ):=S(\rho _{A})+S(\rho _{B})-S(\rho )
\end{equation}%
can be interpreted as the quantum mutual information,
\begin{equation}
S(\rho ):=-\text{Tr}(\rho \log \rho )
\end{equation}%
is the von Neumann entropy,
\begin{eqnarray}
I(\rho |\Pi ^{A}) &:&=S(\rho _{B})-S(\rho |\Pi ^{A})  \notag \\
S(\rho |\Pi ^{A}) &:&=\sum\limits_{k}p_{k}S(\rho _{k})
\end{eqnarray}%
and
\begin{equation}
\rho _{k}=\frac{1}{p_{k}}(\Pi _{k}^{A}\otimes I_{B})\rho (\Pi
_{k}^{A}\otimes I_{B})
\end{equation}%
with
\begin{equation}
p_{k}=\text{Tr}[(\Pi _{k}^{A}\otimes I_{B})\rho (\Pi _{k}^{A}\otimes I_{B})],%
\text{ }k=1,2
\end{equation}%
Measurement-induced nonlocality can be viewed as a kind of quantum
correlation from a geometric perspective based on the local von Neumann
measurements from which one of the reduced states is left invariant. The MIN
of $\rho $, denoted by $MIN(\rho )$, can be defined as [42]%
\begin{equation}
MIN_{A}(\rho ):=\underset{\Pi ^{A}}{\max }\left\Vert \rho -|\Pi ^{A}(\rho
)\right\Vert ^{2}
\end{equation}%
where $\left\Vert .\right\Vert $ stands for the Hilbert-Schmidt norm ($%
\left\Vert A\right\Vert =[Tr(A^{\dagger }A)]^{1/2}$) and the maximum is
taken over all local von Neumann measurement $\Pi ^{A}=\{\Pi _{k}^{A}\}$
with $\sum\limits_{k}\Pi _{k}^{A}\rho _{A}\Pi _{k}^{A}=\rho _{A}$,%
\begin{equation}
\Pi ^{A}(\rho )=\sum\limits_{k}(\Pi _{k}^{A}\otimes I_{B})\rho (\Pi
_{k}^{A}\otimes I_{B}).
\end{equation}%
Whereas the geometric measure of quantum discord for the class of $X$-states
can be defined as [108]

\begin{equation}
D_{G}(\rho ):=\underset{\Pi ^{A}}{\min }\left\Vert \rho -\chi \right\Vert
^{2}
\end{equation}%
where the minimum is over the set of zero-discord states $\chi $. The square
of Hilbert-Schmidt norm of Hermitian operators, $\left\Vert \rho -\chi
\right\Vert ^{2}=$Tr$[(\rho -\chi )^{2}]$. This quantity have been evaluated
by Dakic et al. [6] for an arbitrary two-qubit state.

\section{Results and discussions}

Instead of presenting analytical expressions for the geometric measure of
quantum discord and measurement-induced nonlocality, these quantities are
interpreted in terms of their corresponding graphs for various situations as
illustrated in the figures captions. Influence of different decoherence
channels such as amplitude damping, depolarizing and phase flip channels is
investigated for $X$-type initial states in accelerated frames at finite
temperature. The results consist of three parts (i) the effect of
acceleration $r$ on the GMQD and MIN of $X$-type initial states (ii) the
effect of decoherence on the GMQD and MIN (iii) the effect of finite
temperature $X$ on the GMQD and MIN influenced by the decoherence channels.

In figure 1, the geometric measure of quantum discord (GMQD) and
measurement-induced nonlocality (MIN) are plotted as a function of
acceleration $r$ for Bell, Werner and General type initial states. It is
seen that GMQD can be used to distinguish initial quantum states, whereas in
case of MIN, all the three states overlap. In figure 2, GMQD and MIN are
plotted as a function of decoherence parameter $p$ for Bell, Werner and
General type initial states for $r=0$ (upper panel) and $r=\pi /4$\ (lower
panel) for the case of amplitude damping channel. Here $r=\pi /4$
corresponds to infinite acceleration limit. It can be seen that the
depolarizing channels heavily influences the geometric quantum discord and
MIN as compared to amplitude damping channel. In figure 3, GMQD and MIN are
plotted as a function of decoherence parameter $p$ (upper panel) for Bell,
Werner and General type initial states for $r=\pi /4$ (lower panel) as a
function of decoherence parameter $p$ and acceleration $r$ for Werner type
initial states, for phase flip channel. It is seen that the behaviour of
phase flip channel is symmetrical around 50\% decoherence. It is also seen
that for lower level range of decoherence, different initial states can be
distinguished. Furthermore, GMQD sudden birth appears for higher level of
decoherence for which the initial states become distinguishable. Whereas,
for the intermediate range of decoherence, it is difficult to distinguish
different initial states.

In figure 4, GMQD and MIN are plotted as a function of the parameter $X$ for
General type initial state for $\bar{n}=0.01,$ $\bar{n}=0.1$ and $\bar{n}%
=0.3 $ for $r=0$ (upper panel) and for $r=\pi /4$ (lower panel) \
respectively. It is seen that a sudden change in the behaviour (rise and
then fall) of GMQD occurs depending upon the mean photon number of the local
environment. Whereas, MIN suddenly falls down at finite $\bar{n}$ (i.e. very
small mean photon number) for higher values of the parameter $X$. In figure
5, GMQD is plotted for depolarizing, amplitude damping and phase flip
channels for $\bar{n}=0.01$ (upper graph) and $\bar{n}=0.1$ (lower graph)
for the Werner like states. A sudden change in the behaviour of GMQD is seen
depending upon the choice of the parameter $c$ and the mean photon number $%
\bar{n}$. It is also seen that the mean number of photons plays a crucial
role in the dynamics of GMQD as we increase the value of $\bar{n},$ the
changing behaviour of GMQD saturates for $\bar{n}>0.3.$ The sudden change in
the behaviour of GMQD and MIN become more prominent, when we introduce the
environmental noise in the system.

In figure 6, GMQD (upper panel, with $p=0.5$) and MIN (lower panel, with $%
r=\pi /4$) are plotted as a function of acceleration $r,$ $p$ respectively,
and parameter $X$ for General initial state for $\bar{n}=0.01$ and $\bar{n}%
=0.1$ (a) amplitude damping (b) depolarizing (c) phase flip channels. Here
upper graph corresponds to lower value of the mean \ number of photons of
the local reservoir. It is seen from the figure that as the value of
acceleration $r$ increases, the GMQD and MIN are degraded, the effect is
more prominent in case of phase flip channel. The sudden transition in the
behaviour of GMQD and MIN can be seen for amplitude damping and phase flip
channels. This sudden change is dependent on the mean number of photons of
the local environment. Similarly, sudden change in the behaviour of MIN is
also seen in the finite temperature regime (i.e. at finite $\bar{n}$). It is
noticeable that in case of depolarizing channel, no sudden change in the
behaviour of GMQD and MIN is observed. It means that the depolarizing
channel has destructive interference effect on the local environment.
Furthermore, the environmental noise has stronger affect on the dynamics of
GMQD and MIN as compared to the Unruh effect. It is also seen that Werner
like states are more robust than General type initial states at finite
temperature.

\section{Conclusions}

Decoherence dynamics of geometric measure of quantum discord (GMQD) and
measurement-induced nonlocality (MIN) is investigated for noninertial
observers at finite temperature. A two-qubit $X$-state is considered in
noninertial frames and the effect of acceleration $r$, decoherence and
finite temperature is analyzed. Evolution of GMQD and MIN is studied
influenced by different environments such as amplitude damping, depolarizing
and phase flip channels. It is shown that initial state entanglement plays
an important role in bipartite quantum states. It is possible to distinguish
the Bell diagonal, Werner and general type initial states using the GMQD. A
sudden transition in the behaviour of GMQD and MIN occurs depending upon the
mean photon number of the local environment at finite $\bar{n}$ (i.e. for
small mean photon number). Therefore, mean photon number plays an important
role in the dynamics of GMQD and MIN. The transition behaviour disappears
for larger values of $\bar{n},$ i.e. $\bar{n}>0.3.$ This transition
behaviour become more prominent, when environmental noise is introduced in
the system. In the presence of environmental noise, as we increase the value
of acceleration $r$, GMQD and MIN decay due to Unruh effect which is more
prominent for the phase flip noise. This sudden change is dependent on the
mean number of photons of the local environment. It is notable that for the
depolarizing channel, no sudden change in the behaviour of GMQD and MIN is
observed means that the depolarizing channel has destructive interference
effect on the local environment. Furthermore, the environmental noise has
stronger affect on the dynamics of GMQD and MIN as compared to the Unruh
effect. It is also seen that Werner like states are more robust than General
type initial states at finite temperature.

\section{Appendix-A}

\begin{eqnarray*}
\rho _{11}[t] &=&\frac{1}{^{(2\bar{n}+1)^{2}}}\{\bar{n}^{2}+[2(\rho
_{11}-\rho _{44})\bar{n}^{2}+(\rho _{11}-\rho _{44}+1)\bar{n}]X \\
&&+[(2\rho _{11}+2\rho _{44}-1)\bar{n}^{2}+(3\rho _{11}+\rho _{44}-1)\bar{n}%
+\rho _{11}]X^{2}\} \\
\rho _{22}[t] &=&\frac{1}{^{^{(2\bar{n}+1)^{2}}}}\{\bar{n}(\bar{n}%
+1)-[2(\rho _{11}+2\rho _{33}+\rho _{44}-1)\bar{n}^{2} \\
&&+(\rho _{11}+4\rho _{33}+3\rho _{44}-2)\bar{n}+(\rho _{33}+\rho _{44}-1)]X
\\
&&-[(2\rho _{11}+2\rho _{44}-1)\bar{n}^{2}+(3\rho _{11}+\rho _{44}-1)\bar{n}%
+\rho _{11}]X^{2}\} \\
\rho _{33}[t] &=&\frac{1}{^{^{(2\bar{n}+1)^{2}}}}\{\bar{n}(\bar{n}%
+1)+[2(\rho _{11}+2\rho _{33}+\rho _{44}-1)\bar{n}^{2}+(3\rho _{11}+4\rho
_{33}+\rho _{44}-2)\bar{n}]X \\
&&-[(2\rho _{11}+2\rho _{44}-1)\bar{n}^{2}+(3\rho _{11}+\rho _{44}-1)\bar{n}%
+\rho _{11}]X^{2}\} \\
\rho _{44}[t] &=&\frac{1}{^{^{(2\bar{n}+1)^{2}}}}\{(\bar{n}+1)^{2}-(\bar{n}%
+1)[2\bar{n}(\rho _{11}-\rho _{44})+(\rho _{11}-\rho _{44}-+)]X \\
&&+[(2\rho _{11}+2\rho _{44}-1)\bar{n}^{2}+(3\rho _{11}+\rho _{44}-1)\bar{n}%
+\rho _{11}]X^{2}\} \\
\rho _{14}[t] &=&\rho _{44}X,\text{ }\rho _{23}[t]=\rho _{23}X
\end{eqnarray*}%
where $X=\exp ^{-\Gamma (2\bar{n}+1)t}$ and $\Gamma $\ is the spontaneous
decay rate of the qubits.

{\huge Figures captions}\newline
\textbf{Figure 1}. (Color online). Geometric measure of quantum discord
(GMQD) and measurement-induced nonlocality (MIN) are plotted as a function
of acceleration $r$ for Bell, Werner and General type initial states.\newline
\textbf{Figure 2}. (Color online). GMQD and MIN are plotted as a function of
decoherence parameter $p$ for Bell, Werner and General type initial states
for $r=0$ (upper panel) amplitude damping channel and $r=\pi /4$ (lower
panel) for the case of amplitude damping channel.\newline
\textbf{Figure 3}. (Color online). GMQD and MIN are plotted as a function of
decoherence parameter $p$ (upper panel) for Bell, Werner and General type
initial states for $r=\pi /4$ (lower panel) as a function of decoherence
parameter $p$ and acceleration $r$ for Werner type initial states, for phase
flip channel.\newline
\textbf{Figure 4}. (Color online). GMQD and MIN are plotted as a function of
the parameter $X$ for General type initial state ($c_{1}=0.2,$ $c_{2}=-0.3,$
$c_{3}=0.3$) for $\bar{n}=0.01,$ $\bar{n}=0.1$ and $\bar{n}=0.3$ for $r=0$
(upper panel) and for $r=\pi /4$ (lower panel) \ respectively.\newline
\textbf{Figure 5}. (Color online). GMQD is plotted as a function of
parameter $c$ for depolarizing, amplitude damping and phase flip channels
for $\bar{n}=0.01$ (upper graph) and $\bar{n}=0.1$ (lower graph) for the
Werner like states ($\left\vert c_{1}\right\vert =\left\vert
c_{2}\right\vert =\left\vert c_{3}\right\vert =c$). \newline
\textbf{Figure 6}. (Color online). GMQD (upper panel, with $p=0.5$) and MIN
(lower panel, with $r=\pi /4$) are plotted as a function of acceleration $r,$
$p$ respectively, and parameter $X$ for General initial state ($c_{1}=0.2,$ $%
c_{2}=-0.3,$ $c_{3}=0.3$) with $\bar{n}=0.01$ $($upper graph$)$ and $\bar{n}%
=0.1($lower graph) for amplitude damping, depolarizing and phase flip
channels.\newline
{\Huge Table Caption}\newline
\textbf{Table 1}. Single qubit Kraus operators for amplitude damping,
depolarizing, and phase flip channels where $p$ represents the decoherence
parameter.\newpage

\begin{figure}[tbp]
\begin{center}
\vspace{-2cm} \includegraphics[scale=0.6]{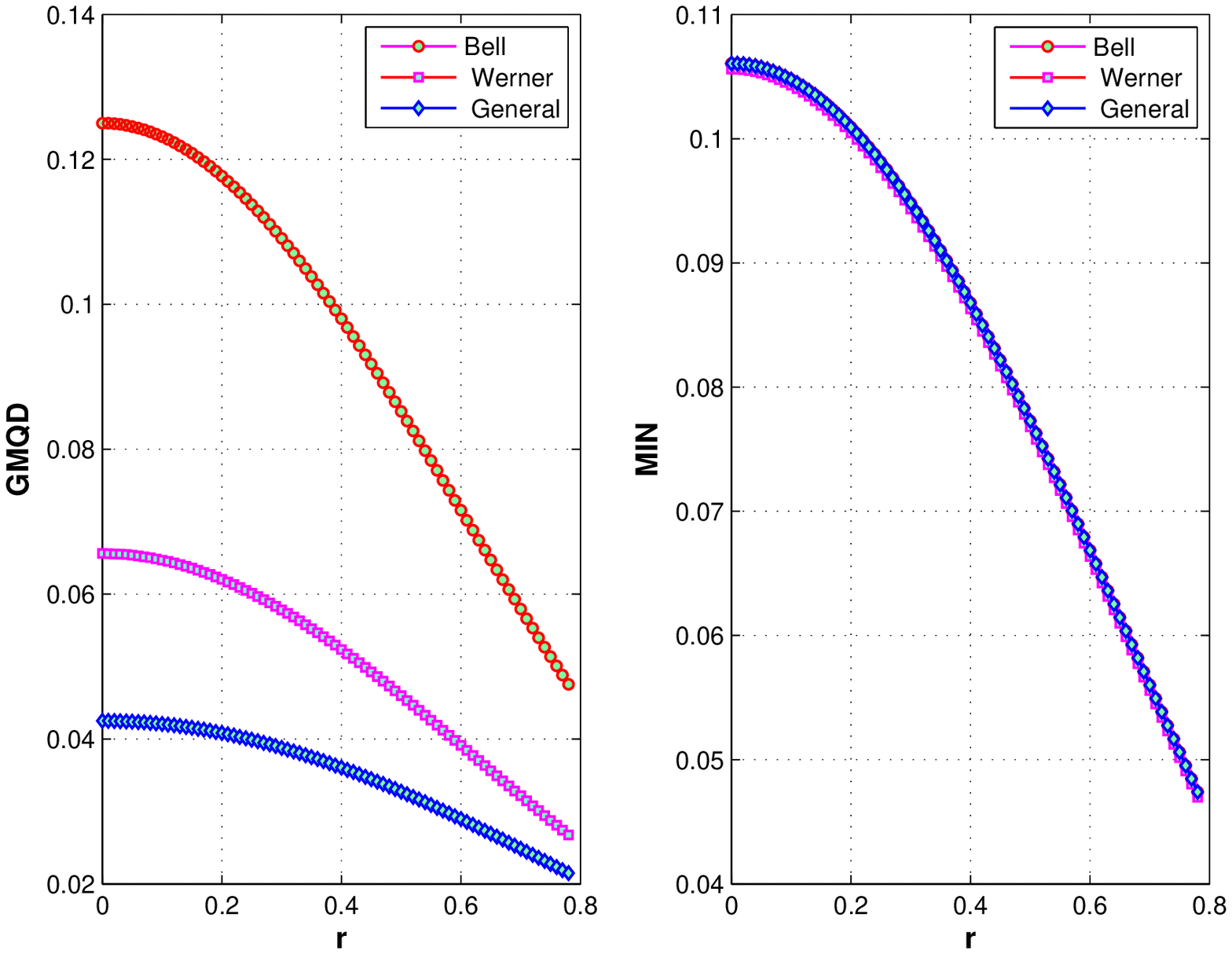} \\[0pt]
\end{center}
\caption{ (Color online). Geometric measure of quantum discord (GMQD) and
measurement-induced nonlocality (MIN) are plotted as a function of
acceleration $r$ for Bell, Werner and General type initial states.\newline
}
\end{figure}
\begin{figure}[tbp]
\begin{center}
\vspace{-2cm} \includegraphics[scale=0.6]{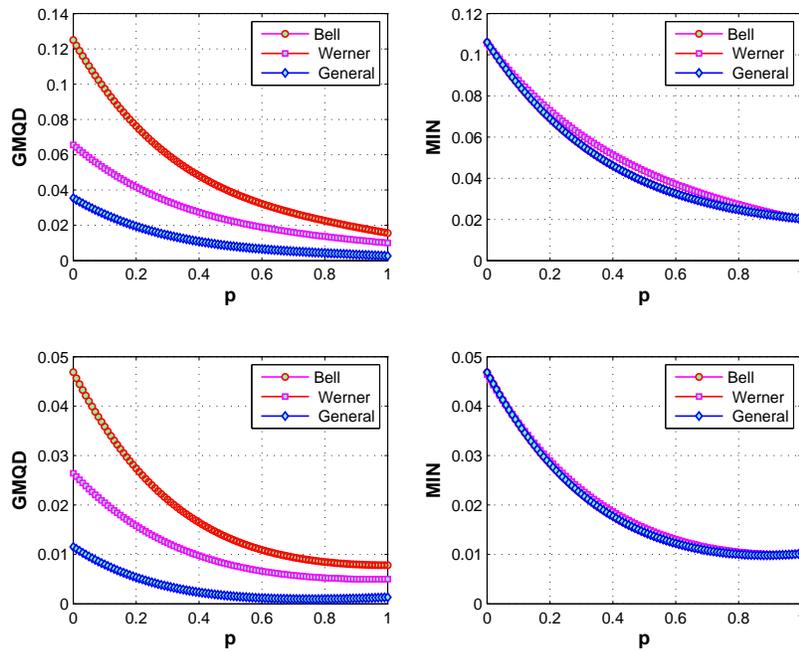} \\[0pt]
\end{center}
\caption{(Color online). GMQD and MIN are plotted as a function of
decoherence parameter $p$ for Bell, Werner and General type initial states
for $r=0$ (upper panel) amplitude damping channel and $r=\protect\pi /4$
(lower panel) for the case of amplitude damping channel.\newline
}
\end{figure}
\begin{figure}[tbp]
\begin{center}
\vspace{-2cm} \includegraphics[scale=0.6]{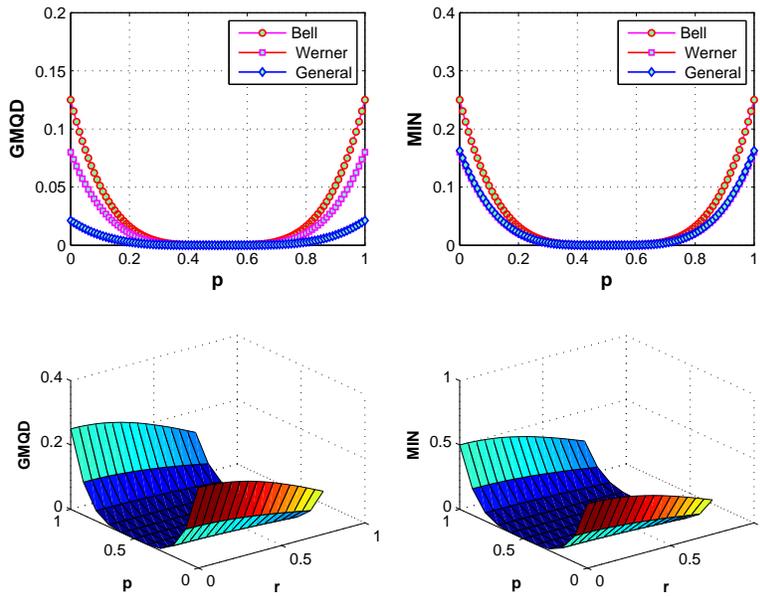} \\[0pt]
\end{center}
\caption{ (Color online). GMQD and MIN are plotted as a function of
decoherence parameter $p$ (upper panel) for Bell, Werner and General type
initial states for $r=\protect\pi /4$ (lower panel) as a function of
decoherence parameter $p$ and acceleration $r$ for Werner type initial
states, for phase flip channel.\newline
}
\end{figure}
\begin{figure}[tbp]
\begin{center}
\vspace{-2cm} \includegraphics[scale=0.6]{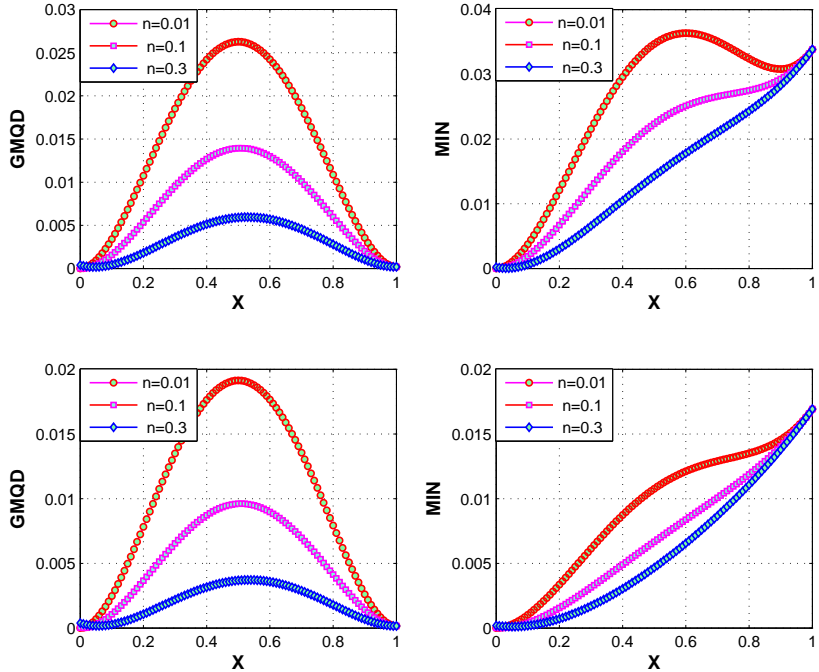} \\[0pt]
\end{center}
\caption{(Color online). GMQD and MIN are plotted as a function of the
parameter $X$ for General type initial state ($c_{1}=0.2,$ $c_{2}=-0.3,$ $%
c_{3}=0.3$) for $\bar{n}=0.01,$ $\bar{n}=0.1$ and $\bar{n}=0.3$ for $r=0$
(upper panel) and for $r=\protect\pi /4$ (lower panel) \ respectively.%
\newline
}
\end{figure}
\begin{figure}[tbp]
\begin{center}
\vspace{-2cm} \includegraphics[scale=0.6]{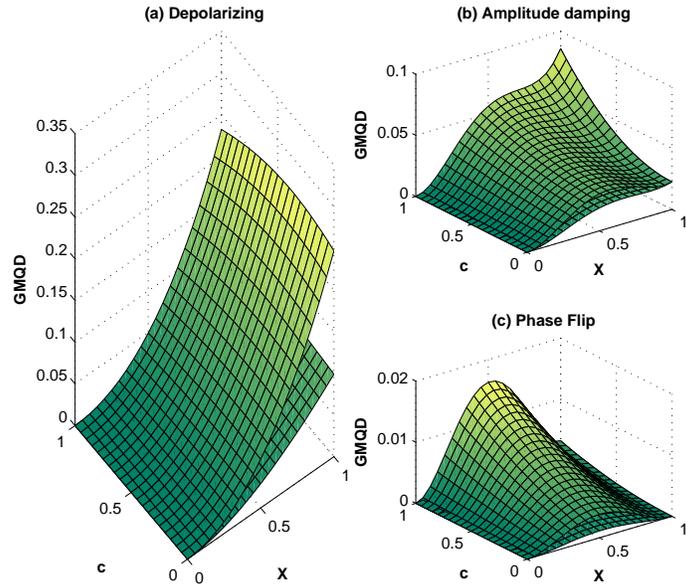} \\[0pt]
\end{center}
\caption{(Color online). GMQD is plotted as a function of parameter $c$ for
depolarizing, amplitude damping and phase flip channels for $\bar{n}=0.01$
(upper graph) and $\bar{n}=0.1$ (lower graph) for the Werner like states ($%
\left\vert c_{1}\right\vert =\left\vert c_{2}\right\vert =\left\vert
c_{3}\right\vert =c$). }
\end{figure}
\begin{figure}[tbp]
\begin{center}
\vspace{-2cm} \includegraphics[scale=0.6]{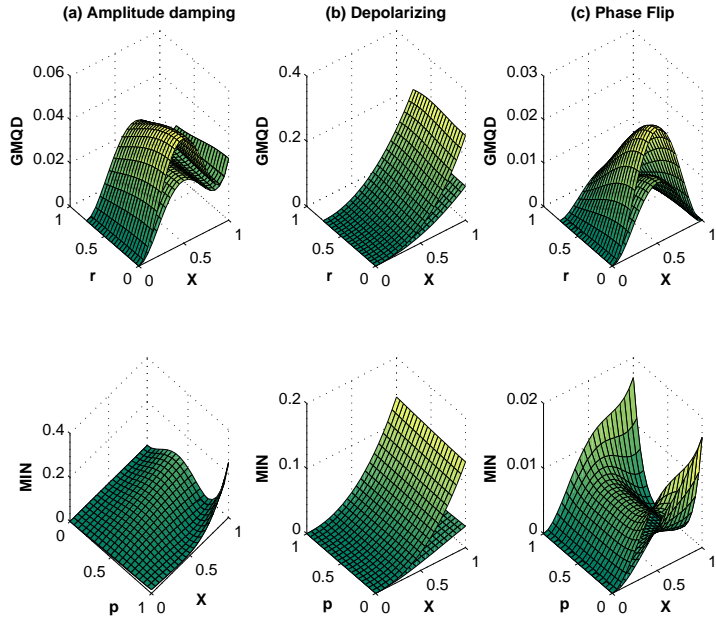} \\[0pt]
\end{center}
\caption{(Color online). GMQD (upper panel, with $p=0.5$) and MIN (lower
panel, with $r=\protect\pi /4$) are plotted as a function of acceleration $%
r, $ $p$ respectively, and parameter $X$ for General initial state ($%
c_{1}=0.2,$ $c_{2}=-0.3,$ $c_{3}=0.3$) with $\bar{n}=0.01$ $($upper graph$)$
and $\bar{n}=0.1($lower graph) for amplitude damping, depolarizing and phase
flip channels.\newline
}
\end{figure}

\begin{table}[tbh]
\caption{Single qubit Kraus operators for amplitude damping, depolarizing,
and phase flip channels where $p$ represents the decoherence parameter.}$%
\begin{tabular}{ll}
&  \\
&  \\ \hline
\multicolumn{1}{|l}{} & \multicolumn{1}{|l|}{} \\
\multicolumn{1}{|l}{$\text{Amplitude damping channel}$} &
\multicolumn{1}{|l|}{$A_{0}=\left[
\begin{array}{cc}
1 & 0 \\
0 & \sqrt{1-p}%
\end{array}%
\right] ,$ $A_{1}=\left[
\begin{array}{cc}
0 & \sqrt{p} \\
0 & 0%
\end{array}%
\right] $} \\
\multicolumn{1}{|l}{} & \multicolumn{1}{|l|}{} \\ \hline
\multicolumn{1}{|l}{} & \multicolumn{1}{|l|}{} \\
\multicolumn{1}{|l}{$\text{Depolarizing channel}$} & \multicolumn{1}{|l|}{$%
\begin{tabular}{l}
$A_{0}=\sqrt{1-\frac{3p}{4}I},\quad A_{1}=\sqrt{\frac{p}{4}}\sigma _{x}$ \\
$A_{2}=\sqrt{\frac{p}{4}}\sigma _{y},\quad \quad $\ $\ A_{3}=\sqrt{\frac{p}{4%
}}\sigma _{z}$%
\end{tabular}%
$} \\
\multicolumn{1}{|l}{} & \multicolumn{1}{|l|}{} \\ \hline
\multicolumn{1}{|l}{} & \multicolumn{1}{|l|}{} \\
\multicolumn{1}{|l}{$\text{Phase flip channel}$} & \multicolumn{1}{|l|}{$%
A_{0}=\sqrt{1-p}I,\quad A_{1}=\sqrt{p}\sigma _{z}$} \\
\multicolumn{1}{|l}{} & \multicolumn{1}{|l|}{} \\ \hline
&  \\
&  \\
&  \\
&
\end{tabular}%
$%
\label{di-fit}
\end{table}

\end{document}